\documentclass[a4paper,11pt]{article}
\pdfoutput=1
\usepackage{jcappub}
\usepackage{longtable}
\usepackage{epsfig}
\usepackage{amssymb}
\usepackage{amsmath}
\usepackage{verbatim}
\usepackage{xspace}
\usepackage{titlesec}
\usepackage{subfigure}
\usepackage{multirow}
\include{jdefs}
\definecolor{darkblue}{rgb}{0,0,0.5}
\definecolor{darkgreen}{rgb}{0.1,0,0.3}
\definecolor{darkred}{rgb}{0.6,0,0}
\usepackage{graphics}
\usepackage{hyperref}

\newcommand{\ba}{\begin{eqnarray}}
\newcommand{\ea}{\end{eqnarray}}
\newcommand{\be}{\begin{equation}}
\newcommand{\ee}{\end{equation}}

\newcommand{\NAG}{\texttt{NAG} }

\usepackage{breqn}
\catcode`_=12
\catcode`^=12
\usepackage{tensor}

\title{\boldmath The gravitational field of a star in quadratic gravity}

\author[a,b]{A. Bonanno}
\author[c,d]{and S. Silveravalle}

\affiliation[a]{INAF, Osservatorio Astrofisico di Catania,\\Via S.Sofia 78, IT-95123 Catania, Italy}
\affiliation[b]{INFN, Sezione di Catania,\\Via S. Sofia 64, IT-95123, Catania, Italy}
\affiliation[c]{Universit\`a degli Studi di Trento,\\Via Sommarive, 14, IT-38123, Trento, Italy}
\affiliation[d]{TIFPA - Trento Institute for Fundamental Physics and Applications,\\Via Sommarive, 14, IT-38123, Trento, Italy}

\emailAdd{alfio.bonanno@inaf.it}
\emailAdd{samuele.silveravalle@unitn.it}

\abstract{The characterization of the gravitational field of isolated objects is still an open question in quadratic theories of gravity. 
We study static equilibrium solutions for a self-gravitating fluid in extensions of General Relativity including 
terms quadratic in the Weyl tensor $C_{\mu\nu\rho\sigma}$ and in the Ricci scalar $R$, as suggested by one-loop corrections to classical gravity. By the means of a shooting method procedure we link the total gravitational mass and the strength of the Yukawa corrections associated with the quadratic terms with the fluid properties at the center.
It is shown that the inclusion of the $C_{\mu\nu\rho\sigma}C^{\mu\nu\rho\sigma}$ coupling in the lagrangian has a much stronger
impact than the $R^2$ correction in the determination of the radius and of the maximum mass of a compact object.
We also suggest that the  ambiguity in the definition of mass in quadratic gravity theories can conveniently 
be exploited to detect deviations from standard General Relativity.}

\begin{document}
\maketitle
\flushbottom

\section{Introduction}
Quadratic theories of gravity naturally emerge as alternative candidates to Einstein's theory at the quantum level. In particular, quadratic terms in the Ricci scalar and Ricci tensor have long been known to appear in the gravitational action as one-loop counterterms of General Relativity \citep{DeWitt:1967ub,tHooft:1974toh}. In more recent times similar terms were also found in the low energy limit of string theories, as for the heterotic $E_8\times E_8$ superstring \citep{Zwiebach:1985uq}, and as marginal-relevant operators in renormalization group approaches to quantum gravity, as in the context of asymptotically safe gravity \citep{Benedetti:2013jk}. In $d=4$ dimensions the most general theory with quadratic combinations of the curvature tensors is defined by the following action
\be\label{action}
\mathcal{S}=\int\mathrm{d}^4x\,\sqrt{-g}\,\left[\gamma\,R-\alpha\,C_{\mu\nu\rho\sigma}C^{\mu\nu\rho\sigma}+\beta\,R^2+\mathcal{L}_{mat}\right],
\ee
where  $C_{\mu\nu\sigma\tau}$ is the Weyl tensor and $\mathcal{L}_{mat}$ is the matter lagrangian.\footnote{This is the most general theory because we can always add a term proportional to the Gauss-Bonnet combination $\mathcal{G}=R^{\mu\nu\rho\sigma}R_{\mu\nu\rho\sigma}-4\,R^{\mu\nu}R_{\mu\nu}+R^2$ that, in $d=4$ dimensions, is a total derivative.} The dynamical content of the theory consists in a spin-two massless Einstein mode and additional massive spin-two and spin-zero modes of masses $m_2^2=\gamma/2\alpha$ and $m_0^2=\gamma/6\beta$ respectively. The theory turns out to be renormalizable \cite{Stelle:1976gc} but the spin-two massive mode is  a ghost excitation which can potentially  spoil the unitarity of the Lorentzian theory. The energy scale of this phenomenon is not known and  various proposals on possible resolutions of this issue have recently appeared \cite{anselmi17,piva,2019PhRvD.100j5006D,Platania:2020knd}. According to Ostrogradsky's theory the presence of these states corresponds  to an instability at the classical level. It is then of great importance to study the classical content of the theory, and in particular the properties of the solutions both in vacuum and in the presence of matter, in order to understand if and how such instabilities affect the physical content of the theory.

In spherical symmetry the weak field limit of asymptotically flat  solutions of (\ref{action}) is well understood at theoretical \cite{Stelle:1977ry} and phenomenological \citep{Accioly:2015fka,Accioly:2016qeb} level. At large $r$  the space-time is in fact completely determined by the $1/r$ newtonian potential  generated by the exchange of the spin-two massless mode and a combination of Yukawa corrections of the type $e^{-m r}/r$, where $m=m_2$ or $m=m_0$ for the spin-two and spin-zero massive modes. In the limit of strong field regime the non-linearity of the field equations renders the study of the spectrum of the possible solutions much more difficult. A major turning point was given by a generalized Israel theorem \citep{Nelson:2010ig,Lu:2015psa} which states that any black hole solution of quadratic gravity has to be a solution of the restricted Einstein-Weyl, i.e. $R+C_{\mu\nu\rho\sigma}C^{\mu\nu\rho\sigma}$, theory. The recent discovery of a class of non-Schwarzschild black holes in this restricted theory \cite{Lu:2015cqa,Lu:2015psa} has triggered new works in this direction. Combinations of analytical \cite{Kokkotas:2017zwt,Podolsky:2018pfe,Konoplya:2020hyk}  and numerical approaches \cite{2018PhRvD..97b4015G,Bonanno:2019rsq,Hernandez-Lorenzo:2020aie} have studied these new  black holes in detail, and in \cite{Bonanno:2019rsq} the first complete analysis of the link between the weak field expansion, the structure of the horizons and the interior has been presented for  static space-times.
Solutions with regular metric coefficients and no horizons as well as  wormholes type solutions have also been discovered \citep{Lu:2015psa,Podolsky:2018pfe,Holdom:2016nek}. All this complex taxonomy  has to be contrasted with the situation in classical  GR where the external field of an asymptotically flat, spherically symmetric space-time is instead uniquely determined by the Schwarzshild metric.

At the same time the success of the Starobinsky model of inflation \citep{Starobinsky:1980te,Akrami:2018odb} has brought the attention of the astrophysical community on the $R+R^2$ theory of gravity, and neutron stars in such theory have been studied in great detail in the last years (see \citep{Arapoglu:2010rz,Orellana:2013gn,Astashenok:2013vza,Yazadjiev:2014cza,Capozziello:2015yza,Resco:2016upv,Astashenok:2017dpo,Sbisa:2019mae} as examples in the literature). Both perturbative and non perturbative methods have been used, together with a large number of different equations of state, nevertheless the behaviour of the external gravitational field of neutron stars in quadratic theories of gravity is still not fully clarified. The conceptual difficulty one immediately faces in this case is that the  standard boundary condition at the stellar surface does not uniquely characterize the external field, as already noticed by \cite{Sbisa:2019mae,Astashenok:2017dpo} for the  $R+R^2$ theory. As we shall see in general a ``measured" astronomical mass, e.g. the mass determined by Kepler's third law in a regime where the Newtonian gravity is reproduced, can be different from the mass obtained from the redshift of the spectral lines as measured by a distant observer. In agreement to what is stated in \citep{Resco:2016upv,Sbisa:2019mae}, and in particular in a similar way as discussed in \citep{Astashenok:2017dpo}, we shall argue that this property can be useful to discriminate between GR and quadratic theories of gravity.

Albeit many extensions of the $R+R^2$ theory have been considered, as more complex $f(R)$ theories \citep{Capozziello:2015yza}, the use of Palatini formalism \citep{Silveira:2021ucz,Herzog:2021wpj}, or $f(\mathcal{G})$ theories \citep{Shamir:2019bcw,Naz:2020mjg}, the inclusion of terms quadratic in the Weyl (or equivalently the Ricci) tensor has attracted relatively little interest \citep{Santos:2011ye,Deliduman:2011nw,Herzog:2021wpj}. Considering the key role played by the Weyl term in black hole solutions of quadratic gravity, we believe that it is crucial to study in detail the effect of such term on self gravitating fluids. As we will see, this term actually has major consequences on the solutions, while the $R^2$ term has only limited impact. In the end, the final goal of this work is to bring together these two lines of research: the more theoretically driven one \citep{Stelle:1977ry,Accioly:2015fka,Accioly:2016qeb,Nelson:2010ig,Lu:2015psa,Lu:2015cqa,Kokkotas:2017zwt,Podolsky:2018pfe,Konoplya:2020hyk,2018PhRvD..97b4015G,Bonanno:2019rsq,Hernandez-Lorenzo:2020aie}, and the more astrophysical and phenomenological one \citep{Arapoglu:2010rz,Orellana:2013gn,Astashenok:2013vza,Yazadjiev:2014cza,Capozziello:2015yza,Resco:2016upv,Astashenok:2017dpo,Sbisa:2019mae,Silveira:2021ucz,Herzog:2021wpj,Shamir:2019bcw,Naz:2020mjg}. At practical level, this reduces to link the explicit Yukawa corrections to the Newtonian potential of \citep{Stelle:1977ry} to a perfect fluid stress-energy tensor, by means of the numerical methods already implemented in \citep{Bonanno:2019rsq}.

The plan of the paper is the following. In section \ref{sec2} the energy momentum tensor and the field equations are presented and their properties in the weak field limit are discussed. 
In section \ref{sec3} the numerical method is explained and a list of possible candidates 
to represent a local ``measure" of the mass-energy in these space-times are discussed.  
Section \ref{sec4} is devoted to the results, with a first part dedicated to the effect of the quadratic terms and a second part dedicated to the different mass definitions, and section \ref{sec5} contains the conclusions. 

\section{Energy-momentum tensor and field equations}\label{sec2}
 
\subsection{General framework and equations}

We consider the equations of motion derived from the minimization of the action (\ref{action})
\begin{dmath}\label{eom}
\mathcal{G}_{\mu\nu}=\gamma \left(R_{\mu\nu}-\frac{1}{2}Rg_{\mu\nu}\right)-4\alpha\left(\nabla^\rho\nabla^\sigma +\frac{1}{2}R^{\rho\sigma} \right)C_{\mu\rho\nu\sigma}+2\beta\left(R_{\mu\nu}-\frac{1}{4}Rg_{\mu\nu}-\nabla_\mu\nabla_\nu +g_{\mu\nu}\Box\right)R=\frac{1}{2}T_{\mu\nu},
\end{dmath}
and the static and spherically symmetric ansatz for the metric
\be\label{metric}
ds^2=-h(r)\,dt^2+\frac{dr^2}{f(r)}+r^2d\Omega^2.
\ee
We take the stress-energy tensor with the standard perfect fluid form
\be\label{set}
T_{\mu\nu}=\big(\rho(r)+p(r)\big)\,u_\mu u_\nu+p(r)\,g_{\mu\nu},
\ee
with $u^\mu$ a unit timelike vector. With the ansatz (\ref{metric}) there are only two independent equations of motion, and is possible to reduce the system to third order \citep{Lu:2015psa}.
\be\label{eom3}
\begin{split}
\mathcal{G}_{rr}=&\frac{1}{2}T_{rr},\\
\mathcal{G}_{tt}+X(r)\partial_r\left(\mathcal{G}_{rr}-\frac{1}{2}T_{rr}\right) +Y(r)\left(\mathcal{G}_{rr}-\frac{1}{2}T_{rr}\right)=&\frac{1}{2}T_{tt},
\end{split}
\ee 
where $X(r)$ and $Y(r)$ are combinations of $\alpha$, $\beta$, $r$, $h(r)$, $f(r)$, $\rho(r)$, $p(r)$ and their first and second derivatives, are indeed third order equations respectively in $h(r)$ and $f(r)$. The conservation of the stress-energy tensor results in 
\be\label{conset}
p' (r)=-\frac{h' (r)}{2\,h(r)}\big(\rho(r)+p(r)\big),
\ee
and with the addition of an equation of state
\be\label{geneos}
p(r)=\mathcal{P}\big(\rho(r)\big)
\ee
we have a full set of equation in the variables $h(r)$, $f(r)$, $\rho(r)$ and $p(r)$, that are the homologous of the TOV equations of General Relativity. The full form of the equations (\ref{eom3}) is cumbersome and not very instructive (see Appendix \ref{appendixA}), we note however that in the limit $\alpha,\beta\to 0$ the e.o.m. tensor $\mathcal{G}_{\mu\nu}$ becomes the Einstein tensor $G_{\mu\nu}$, and the combination
\begin{equation}
X(r)\partial_r\left(\mathcal{G}_{rr}-\frac{1}{2}T_{rr}\right) +Y(r)\left(\mathcal{G}_{rr}-\frac{1}{2}T_{rr}\right)
\end{equation}
equals zero whenever the first equation of (\ref{eom3}) and equation (\ref{conset}) are taken into consideration, thus recovering the GR limit.

\subsection{Solutions in the weak field limit}
The field equations (\ref{eom}) can be solved analytically in the weak field limit.
It is convenient to write
\begin{equation}
h(r)=1+\epsilon\,V(r),\qquad f(r)=1+\epsilon\,W(r),
\end{equation}
and expand (\ref{eom}) at linear order in $\epsilon$.  The resulting equations are the 
following
\begin{equation}\label{eoml}
\begin{split}
\tensor{\mathcal{G}}{^\mu_\mu}&=-\left(6\beta\nabla^2-\gamma\right)\left( \nabla^2 V(r)+2Y(r)\right)=\frac{1}{2}\left(-\rho(r)+3p(r)\right)=\frac{1}{2}\tensor{T}{^\mu_\mu},\\
\tensor{\mathcal{G}}{^i_i}-\tensor{\mathcal{G}}{^t_t}&=-4 \left( \beta -\frac{1}{3}\alpha\right ) \nabla^2 Y(r) -2 \bigg ( \beta + \frac{2}{3}\alpha \bigg ) \nabla^2 \nabla^2 V(r) + \gamma \nabla^2 V(r)=\frac{1}{2}\left(\rho(r)+3p(r)\right)\\
&=\frac{1}{2}\left(\tensor{T}{^i_i}-\tensor{T}{^t_t}\right),
\end{split}
\end{equation}
where $Y(r)=r^{-2}\left(rW(r)\right)'$ and $\epsilon$ is set to 1 after the expansion, as described in \citep{Stelle:1977ry,Lu:2015psa,Bonanno:2019rsq}. 
In the vacuum case equations (\ref{eoml}) can be solved by Fourier modes, and we obtain  
\begin{equation}\label{wflbis}
\begin{split}
h(r)=&1+C_T-\frac{2M}{r}+2S_2^+\frac{\mathrm{e}^{m_2r}}{r}+2S_2^-\frac{\mathrm{e}^{-m_2r}}{r}+S_0^+\frac{\mathrm{e}^{m_0r}}{r}+S_0^-\frac{\mathrm{e}^{-m_0r}}{r},\\
f(r)=&1-\frac{2M}{r}+S_2^+\frac{\mathrm{e}^{m_2r}}{r}(1-m_2\,r)+S_2^-\frac{\mathrm{e}^{-m_2r}}{r}(1+m_2\,r)-S_0^+\frac{\mathrm{e}^{m_0r}}{r}(1-m_0\,r)\,+\\
&-S_0^-\frac{\mathrm{e}^{-m_0r}}{r}(1+m_0\,r),\\
\end{split}
\end{equation}
where $m_2^2=\gamma/2\alpha$ and $m_0^2=\gamma/6\beta$. Imposing asymptotic flatness and fixing a parametrization of the time coordinate (i.e. $S_2^+=S_0^+=C_T=0$) we are with the weak field solution
\begin{equation}\label{wfl}
\begin{split}
h(r)=&1-\frac{2M}{r}+2S_2^-\frac{\mathrm{e}^{-m_2r}}{r}+S_0^-\frac{\mathrm{e}^{-m_0r}}{r},\\
f(r)=&1-\frac{2M}{r}+S_2^-\frac{\mathrm{e}^{-m_2r}}{r}(1+m_2\,r)-S_0^-\frac{\mathrm{e}^{-m_0r}}{r}(1+m_0\,r),\\
\end{split}
\end{equation}
that is the Schwarzschild metric with the addition of exponentially suppressed corrections. We note that the weak field solutions of the $R+R^2$ and $R+C_{\mu\nu\rho\sigma}C^{\mu\nu\rho\sigma}$ theories are easily found taking the limits $m_2\to\infty$ and $m_0\to\infty$ respectively.
The equations (\ref{eoml}) can be solved with similar methods also in the presence of a non zero stress-energy tensor and, outside the star, still result in (\ref{wfl}) with a dependence of the free parameters from the pressure and density as
\begin{equation}\label{param}
\begin{split}
M&=\frac{1}{16\pi\gamma}\int_0^{\infty}\mathrm{d}s\,4\,\pi\,s^2\rho(s),\\
S_2^-&=\frac{1}{16\pi\gamma}\int_0^{\infty}\mathrm{d}s\,4\,\pi\,s^2\frac{2\sinh\left(m_2\,s\right)}{3\,m_2\,s}\left(2\,\rho(s)+3\,p(s)\right),\\
S_0^-&=\frac{1}{16\pi\gamma}\int_0^{\infty}\mathrm{d}s\,4\,\pi\,s^2\frac{2\sinh\left(m_0\,s\right)}{3\,m_0\,s}\left(-\rho(s)+3\,p(s)\right),
\end{split}
\end{equation}
in agreement with the result of \citep{Stelle:1977ry} if $\rho(r)=M\delta^3(\vec{x})$ 
and $p(r)=0$.  
The relations (\ref{param}) clearly show that also in the $p=0$ case, the energy-density 
of the matter component determines not only $M$ but also 
the Yukawa coefficients $S^{-}_0$ and $S^{-}_2$. It is therefore important to extend 
this analysis to the fully non-linear regime in order to study the physical implications
of this result. 

\section{Numerical methods and mass definition}\label{sec3}
\subsection{The shooting method and the equation of state}
The interior and exterior field configuration describing a 
compact stars can conveniently be obtained by means of the shooting method. We assume the weak field limit (\ref{wfl}) to be valid at large distances, and integrate the e.o.m. 
with zero energy density and pressure to a radius $r=R_*$. 
At the same time we consider a regular power expansion
\begin{equation}\label{expa}
\begin{split}
h(r)&=h_0(1+h_1\,r+h_2\,r^2+...),\\
f(r)&=f_0+f_1\,r+f_2\,r^2+...,\\
\rho(r)&=\rho_0+\rho_1\,r+\rho_2\,r^2+...,\\
p(r)&=p_0+p_1\,r+p_2\,r^2+...,
\end{split}
\end{equation}
with free parameters $h_0$, $h_2$, $f_2$, $\rho_0$ close to the origin. We then integrate the e.o.m. (\ref{eom3}-\ref{geneos}) to the same radius $r=R_*$ using (\ref{expa}) at fifth order as boundary conditions. To integrate the equations we used the Adaptive Stepsize Runge-Kutta integrator DO2PDF implemented by the \NAG group (see \url{https://www.nag.com} for details) with a tolerance of $10^{-9}$. At radius $R_*$ we obtain continuity of $h(r)$, $f(r)$, their first and second derivatives, $\rho(r)$ and $p(r)$ as function of $(M,S_2^-,S_0^-,h_0,h_2,f_2,\rho_0)$ by means of a globally convergent Newton-Raphson method, as described in \url{http://numerical.recipes}, with a tolerance of $10^{-6}$, leaving only one free parameter. The radius $R_*$ is then defined from the boundary condition $\rho(r=R_\star)=0$ as the surface of the star.
As large radius we chose the value $r_\infty = 18$, in order to have Yukawa corrections larger than the tolerance threshold, while as small radius we used $r_0 = 10^{-3}$ in order to have the discarded terms of the series (\ref{expa}) smaller than the tolerance threshold. To study the GR, $R+R^2$ and $R+C_{\mu\nu\rho\sigma}C^{\mu\nu\rho\sigma}$ cases we used exactly the same method, with the only difference being in the number of free parameters (respectively 3, 5 and 5) due to the lower differential order of the equations.
As equation of state to model the inside of neutron stars we opted for two different choices: a simple polytropic model
\begin{equation}\label{polytrope}
p=k\,\rho^\Gamma
\end{equation}
with $\Gamma=2$ and $k=6.51185\cdot 10^{-17} cm^3/g$, and a more realistic SLy equation of state \citep{Chabanat:1997un,Douchin:2001sv} in its analytical representation \citep{Haensel:2004nu}
\begin{equation}\label{sly}
\begin{split}
\log_{10}p=&\frac{a_1+a_2\log_{10}\rho+a_3\left(\log_{10}\rho\right)^3}{\mathrm{exp}\left[a_5\left(\log_{10}\rho-a_6\right)\right]+1}+\frac{a_7+a_8\log_{10}\rho}{\mathrm{exp}\left[a_9\left(a_{10}-\log_{10}\rho\right)\right]+1}+\\
&+\frac{a_{11}+a_{12}\log_{10}\rho}{\mathrm{exp}\left[a_{13}\left(a_{14}-\log_{10}\rho\right)\right]+1}+\frac{a_{15}+a_{16}\log_{10}\rho}{\mathrm{exp}\left[a_{17}\left(a_{17}-\log_{10}\rho\right)\right]+1},
\end{split}
\end{equation}
where the pressure and density are expressed respectively in $dyn/cm^2$ and $g/cm^3$, and the $a_i$ are numerical parameters that can be found in \citep{Haensel:2004nu}. The polytropic e.o.s. is able to capture the qualitative behaviour of solutions and, being more easily integrated, has been used to show the dependence of the solutions from the action and from the relative values of the masses $m_0$ and $m_2$ in the full theory. The Sly e.o.s., on the other hand, is a good description of what is supposed to be the interior of a neutron star, and has been used in the discussion on mass definition in order to have realistic mass-radius relations.

\subsection{Notes on adimensionalization and the scales of the solutions}\label{sec3su2}
The masses $m_2$ and $m_0$ in (\ref{wfl}) naturally introduce energy and length scales for the solutions. In the restricted $R+R^2$ and $R+C_{\mu\nu\rho\sigma}C^{\mu\nu\rho\sigma}$ theories we choose the respective particle mass (together with $\hbar=c=1$) as natural unit for the adimensionalization of the equations; in the full quadratic theory we choose the mass $m_2$, introducing the parameter $\xi=m_0/m_2$. The choice has been made in order to have a better comparison between the results of quadratic gravity and the ones of the Einstein-Weyl theory \citep{Lu:2015psa,Bonanno:2019rsq,Lu:2015cqa}, in particular with the non-Schwarzschild black holes present in the latter case. A naturalness principle suggests values for $\xi$ of order unity, and the presence of $\xi$ in the exponentials of (\ref{wfl}) forces us to choose values such that
%\begin{equation}
$\frac{\mathrm{e}^{-\xi\,r_\infty}}{r_\infty}>tol, $
%\end{equation}
where $tol$ is the tolerance threshold of the Runge-Kutta integrator.
For these reasons we opted to investigate $\xi$ in the range $\left[0.5-1.5\right]$.\\[0.2cm]
The analytical representation (\ref{sly}) is consistent only in the range of energy densities of order $\left[10^4-10^{16}\right]g/cm^3$, and it is clear that this sets a constraint on the scales of the solutions that can be described using such equation of state. In particular in the general quadratic theory we have that the dimensionful and dimensionless energy densities relate as
\begin{equation}
\rho_{df}\simeq \rho_{dl}\frac{10^{90}}{\alpha} g/cm^3.
\end{equation}
Having the minimum dimensionless value fixed at $10^{-6}$ by the tolerance threshold of the root-finding algorithm, we have to choose values of $\alpha$ not greater than $10^{80}$. We believe that, having found most of the dimensionless values of the density in the range $\left[10^{-6}-10^0\right]$, a value of $\alpha$ of order $10^{74}$ might be optimal. For the evaluation of (\ref{sly}), and while showing the results, we will restore physical units fixing the length scale $l_2=1/m_2$ equal to the Sun Schwarzschild radius $r_{s,\odot}=2\,G\,M_\odot$, in a similar, yet different, fashion to what have been done in \citep{Sbisa:2019mae}. We use the same units, i.e. the unit length equals to the Sun Schwarzschild radius, also for the $R+R^2$ and GR results, in order to have an explicit comparison with the $R+C_{\mu\nu\rho\sigma}C^{\mu\nu\rho\sigma}$ and full quadratic cases. The numerical value for the parameter $\alpha$ (or $\beta$) is then actually of order $10^{74}$, as required by the SLy equation of state. This value exceed the $10^{60}$ laboratory limit obtained for the  Yukawa correction to the gravitational potential \citep{Kapner:2006si,Giacchini:2016nta}. We would like to underline that such constraint directly applies only to the $R+R^2$ and $R+C_{\mu\nu\rho\sigma}C^{\mu\nu\rho\sigma}$ theories, where only a single Yukawa correction is present, as in the case considered by \citep{Kapner:2006si,Giacchini:2016nta}. Nevertheless, although the above limits do not directly apply to the general quadratic theory, we stress that the purpose of our paper is not to provide new constraints on possible modifications of the gravitational potential, rather to discuss the astrophysical properties of the compact star solutions of quadratic gravity.

\subsection{Quasi local masses}\label{massdef}
The results of the weak-field limit showed that the coefficient of the  $1/r$  component of the metric does not uniquely characterize the metric in these space-times. It is expected that in the strong field regime the contribution from the Yukawa terms significantly contribute to  the mass-energy distribution. In these cases a quasi-local definition can be more appropriate to describe the physical phenomena, as already noticed in \citep{Resco:2016upv,Sbisa:2019mae}.
According to this idea one associates a quasi-local  mass-energy to a topologically 2-surface embedded in an asymptotically flat space-time.   Various definitions  (in principle there could be infinite) 
have appeared in the literature in classical general relativity \citep{1968JMP.....9..598H,Hayward:1993ph} 
as the requirement to be consistent with the ``global" mass-energy of an asymptotically flat
stationary space-time defined by the  ADM mass \citep{Arnowitt:1962hi} leaves much freedom in this respect. 
This latter is defined in terms of a 3+1 foliation of an asymptotically flat space-time and it is usually written as
\begin{equation}
E=\gamma\lim_{r\to\infty}\sum_{i,j}\int_{S^2(r)} \mathrm{d}A\, n_i\left(\partial_j g_{ij}-\partial_i g_{jj}\right),
\end{equation}
with $n_i$ the unit normal vector to the sphere $S^2(r)$. In our case we can safely use the weak field limit and, having exponentially suppressed corrections to the Schwarzschild solution, it is clear that results in the $M$ parameter of (\ref{wfl}).  
Useful definitions of quasi-local mass are instead the following.
\paragraph{Misner-Sharp mass} 
Given a spherically symmetric space-time with metric 
\begin{equation}
ds^2= g_{ab}dx^a dx^b+r^2 d\Omega^2
\end{equation}
where $g_{ab}$ is the induced metric in the effective 1+1 space-time, the Misner-Sharp mass
is defined as
\begin{equation}\label{mhay}
 r_{,a} r^{,a}=f(r) =:  1-\frac{2 M(r)}{r},
\end{equation}
so that
\begin{equation}
M_{M{\text -}S}(r)=\frac{1}{2}r(1-f(r))
\end{equation}
in our case. 
It was originally proposed by Misner and Sharp in the context of spherically symmetric models of gravitational collapse \cite{1964PhRv..136..571M}
and it represents the mass-energy enclosed in a spherical hypersurface at time $t$.
It has been generalized by Hayward \citep{Hayward:1993ph} beyond spherical symmetry
and it is widely used in quadratic theories of gravity \citep{Arapoglu:2010rz,Orellana:2013gn,Astashenok:2013vza,Yazadjiev:2014cza,Deliduman:2011nw}.

\paragraph{TOV mass} In General Relativity  
the total mass-energy inside a spherical distribution 
of matter is obtained from the 00-component of the field equations, that with the Misner-Sharp mass definition and our conventions reads
\begin{equation}
M'(r)=\frac{1}{4\gamma} r^2\,\rho(r),
\end{equation}
and which implies
\begin{equation}\label{mtov}
M(r)=\int_0^{r}\mathrm{d}s\, 4\pi s^2 \rho(s).
\end{equation}
The integration extends to the surface $r=R_\ast$ defined by $\rho(R_\ast)=0$
in the Tolman-Oppenheimer-Volkoff equations (TOV) for the relativistic stellar structure \citep{Tolman:1939jz,Oppenheimer:1939ne}, and reaches the limit value $M_{TOV}$.
As it is well known, $M_{TOV}$ does not coincides with the 
proper mass inside the star, but fully describes the observational properties of 
a compact object (neutron star) in General Relativity. 
This definition is strongly dependent from the equation of motion of GR, however, being sensible in the weak field regime (\ref{param}), it has been used also in quadratic theories of gravity \citep{Santos:2011ye,Capozziello:2015yza}. 

\paragraph{Particle potential mass} In order to define the mass of a star in terms of observational properties, Resco et al. proposed a definition based on the effective potential of a massive particle \citep{Resco:2016upv}. With our ansatz the particle equation of motion reads
\begin{equation}\label{pp}
\frac{1}{2}m\,\dot{r}^2+\frac{m\,f(r)L^2}{2\,r^2}+\frac{m}{2}\left(h(r)-1\right)\frac{f(r)}{h(r)}=\frac{f(r)}{h(r)}\left(E-m\right).
\end{equation}
Comparing the third term on the left hand side of (\ref{pp}) with the Newtonian (or General Relativistic) potential energy $-\frac{m\,M}{r}$ we can define the mass as
\begin{equation}\label{mpot}
M_{Pot}(r)=\frac{1}{2}r\left(1-h(r)\right)\frac{f(r)}{h(r)}.
\end{equation}

\paragraph{Newtonian limit mass} In the usual non-relativistic limit of General Relativity the gravitational potential is expressed as
\begin{equation}\label{newlim}
\phi(r)=\frac{1}{2}\left(h(r)-1\right);
\end{equation}
equating (\ref{newlim}) to the Newtonian potential $\phi(r)=-\frac{M}{r}$ we can define the Newtonian limit mass as
\begin{equation}\label{mnew}
M_{New}(r)=\frac{1}{2}r\left(1-h(r)\right).
\end{equation}
The motivation behind this definition is conceptually very similar to the one of Resco et al. (\ref{mpot}), and in fact it could be derived from the particle equation of motion (\ref{pp}) imposing that the energy on the right hand side is independent from the radial coordinate.\\
This definition, however, is particularly useful, being dependent only form the time component of the metric and then associated with the redshift of a photon emitted at radius $r$ and measured at infinity
\begin{equation}
z(r)=\frac{1-\sqrt{h(r)}}{\sqrt{h(r)}}, 
\end{equation}
and therefore it has already been used in modified theories of gravity \citep{Astashenok:2017dpo}.

\paragraph{Kepler's law mass} Another measurable mass definition is the one that can be inferred from the orbital period using Kepler's third Law. Assuming that in some limit
the Newtonian regime is recovered, one can define the Keplerian mass from a measure
of orbital period $T$
\begin{equation}\label{kl}
\frac{r^3}{T^2}=\frac{M}{4\pi^2}.
\end{equation}
The radial geodesic equation, in the case of a circular orbit on the equatorial plane, can be written as
\begin{equation}
\frac{1}{2}h'(r)\,\dot{t}^2-r\,\dot{\phi}^2=0\quad\implies\quad \left(\frac{\mathrm{d}\phi}{\mathrm{d}t}\right)^2=\frac{1}{2}\frac{h'(r)}{r},
\end{equation}
and combined with (\ref{kl}) give rise to the definition
\begin{equation}\label{mkep}
M_{Kep}(r)=\frac{1}{2}r^2h'(r).
\end{equation}

\paragraph{Komar quasi-local mass}
As it is well known 
in a static space-time it is always possible to define the mass in a natural way as the conserved quantity associated to the timelike Killing vector
\begin{equation}\label{mw}
M=-4\gamma\int_S\mathrm{d}A\,n_\mu\frac{\kappa^\nu\nabla_\nu\kappa^\mu}{\sqrt{-\kappa_\rho\kappa^\rho}},
\end{equation}
where $\kappa^\mu$ is the timelike Killing vector, $S$ is a two dimensional surface and $n^\mu$ is the unit normal to such surface \citep{Wald:1984rg}. In General Relativity it can be proved that in the vacuum the integral in (\ref{mw}) is independent from the choice of $S$, and the definition is well cast, while in quadratic gravity this is possible only in the asymptotically infinite region. We can however exploit the definition in (\ref{mw}) and consider it as the energy inside the surface $S$; as usual with our ansatz and units it becomes
\begin{equation}\label{mkom}
M_{Kom}(r)=\frac{1}{2}r^2\sqrt{\frac{f(r)}{h(r)}}h'(r).
\end{equation}

\section{Numerical results}\label{sec4}

\subsection{Stellar structure and global properties in quadratic gravity}

\subsubsection{Single solution: stellar structure and gravitational field}
\begin{figure}[t]
\centering
\includegraphics[width=\textwidth]{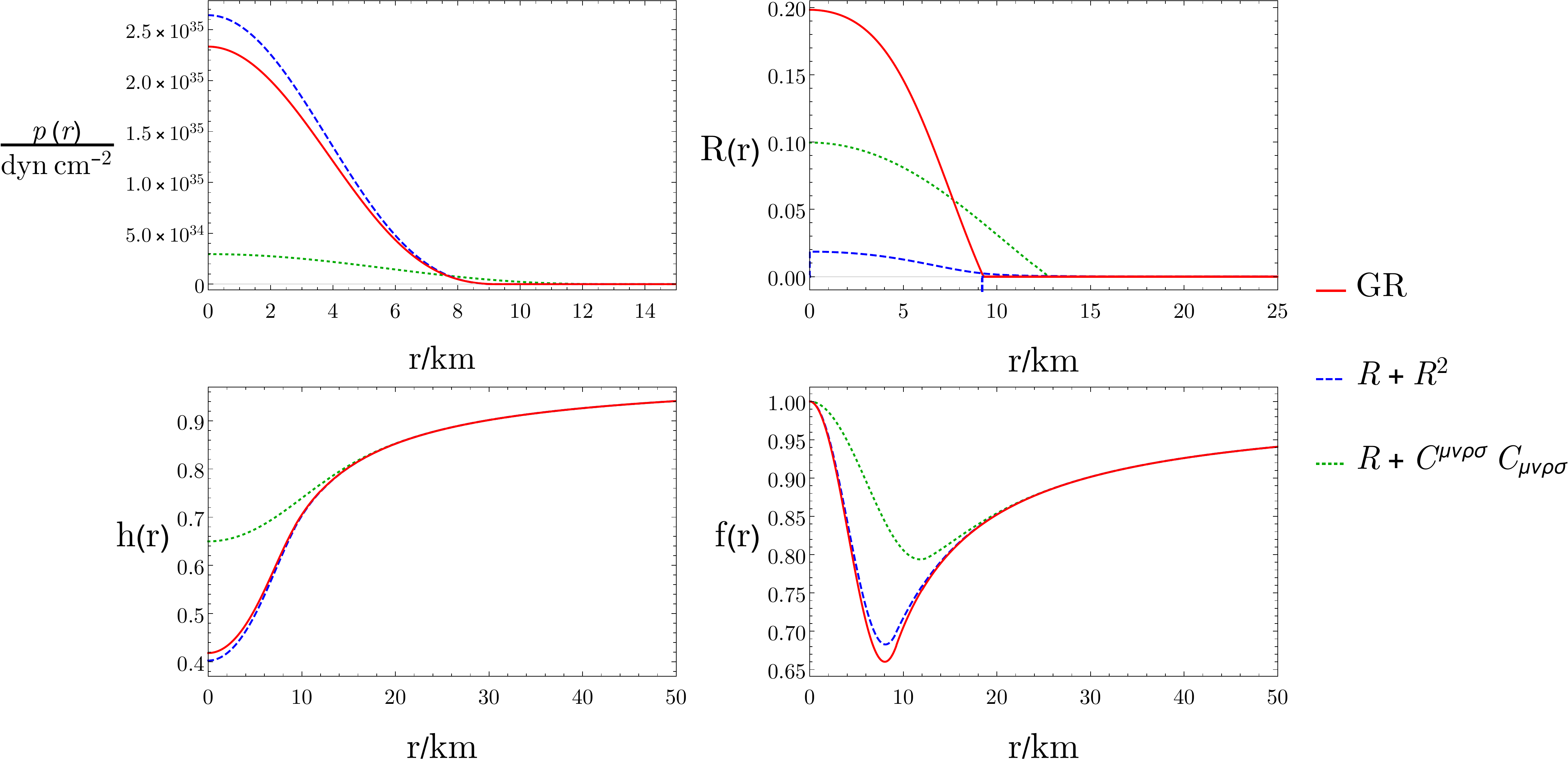}
\caption{Structure of three $M=M_{\odot}$ stars for GR, $R+R^2$ and $R+C_{\mu\nu\rho\sigma}C^{\mu\nu\rho\sigma}$ theories; the dashed vertical lines in the top-right panel indicate the star surface in the $R+R^2$ case. From left to right and top to bottom are: pressure, Ricci scalar, temporal component of the metric and radial component of the metric.}
\label{structureactions}
\end{figure}
\begin{figure}[t]
\centering
\includegraphics[width=\textwidth]{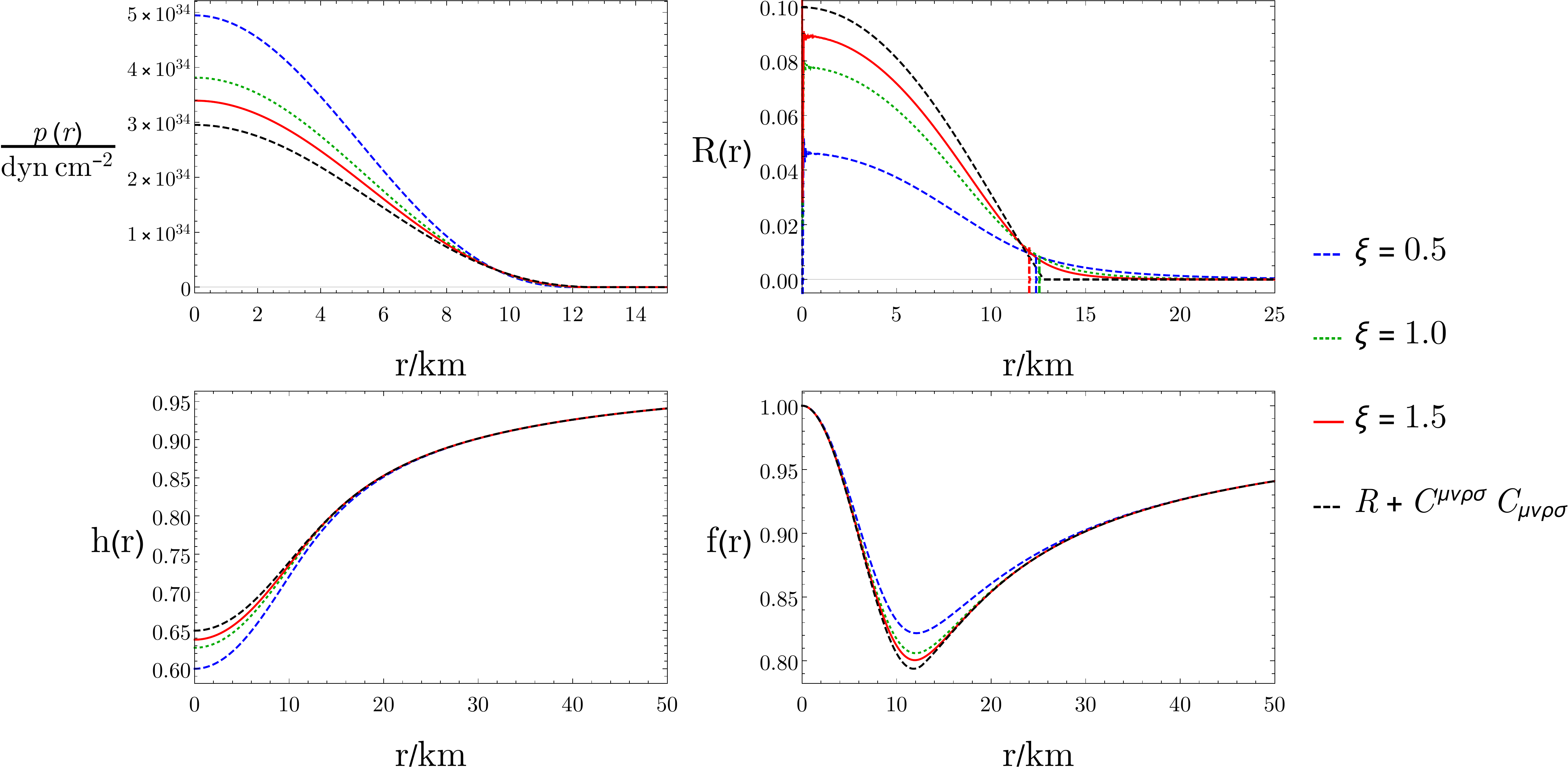}
\caption{Structure of three $M=M_{\odot}$ stars with different values of $\xi$, and the reference $R+C_{\mu\nu\rho\sigma}C^{\mu\nu\rho\sigma}$ solution; the dashed vertical lines in the top-right panel indicate the star surfaces in the full quadratic cases. From left to right and top to bottom are: pressure, Ricci scalar, temporal component of the metric and radial component of the metric.}
\label{structure}
\end{figure}

A first insight on the effect of the two massive modes can be drawn from explicit star structures. We present in figure \ref{structureactions} the behaviour of the metric, scalar curvature and pressure for three stars with the same mass in the restricted general relativistic, $R+R^2$ and $R+C_{\mu\nu\rho\sigma}C^{\mu\nu\rho\sigma}$ cases; in figure \ref{structure}, instead, we show the same quantities for three stars with the same mass in the full quadratic theory, but with different values of $\xi=m_0/m_2$. From figure \ref{structureactions} it seems that the presence of the massive scalar results in a slight strengthening of gravity, with a deeper non relativistic gravitational potential $\phi(r)\propto \left(h(r)-1\right)$ and higher internal pressure. The massive spin-2 particle, on the contrary, leads to a major softening of the gravitational interaction, with much smaller internal pressure and non relativistic gravitational potential. On the other hand, the inclusion of both terms results in a decrease in the radial distortion and in the scalar curvature. If in the Einstein-Weyl case we could relate this effect to the decrease in the fluid pressure and energy density, in the $R+R^2$ case we have to connect this behaviour with a specific property of the theory.

From figure \ref{structure} we see that the full quadratic case is mainly affected by the presence of the Weyl term, with a global softening of gravity, that is less pronounced as the value of $\xi$ is lowered. Here we have to remember that lower values of $\xi$ mean that the $R^2$ term is dominant with respect to the $C_{\mu\nu\rho\sigma}C^{\mu\nu\rho\sigma}$ term in the action (\ref{action}), or, from a particle point of view, the range of the massive scalar contribution to the gravitational interaction is larger than the massive tensorial one. It is also more clear the impact of the $R^2$ term on the scalar curvature, which is to increase the curvature outside the star and to decrease it inside. The $R^2$ term also contribute to smooth even more the radial component of the metric, decreasing the radial distortion. The behaviour of the Ricci scalar is better understood looking at the trace of the equation of motion (\ref{eom})
\begin{equation}\label{eomtrace}
\left(6\beta\,\Box-2\gamma\right)R=\frac{1}{2}T\qquad\implies\qquad \left(\frac{1}{\xi^2}\Box-2\right)R=\frac{1}{2}T,
\end{equation}
where $T=\tensor{T}{^\mu_\mu}$ and on the right we used our units. For higher values of $\xi$ the Ricci scalar gets closer to its form in General Relativity $R\propto T$, which is zero outside and large inside the star. For lower values of $\xi$, instead, we have that the presence of the fluid has little effect on scalar curvature that, having imposed regularity in the origin and asymptotic flatness, flattens out and gets closer to zero. Having a deeper gravitational potential, the flattening of the scalar curvature has to be driven by a more regular behaviour of $f(r)$, or in other words, by a decrease in the radial distortion.

\subsubsection{Families of solutions: parameters relations and physical interpretation}

\begin{figure}[t]
\centering
\includegraphics[width=\textwidth]{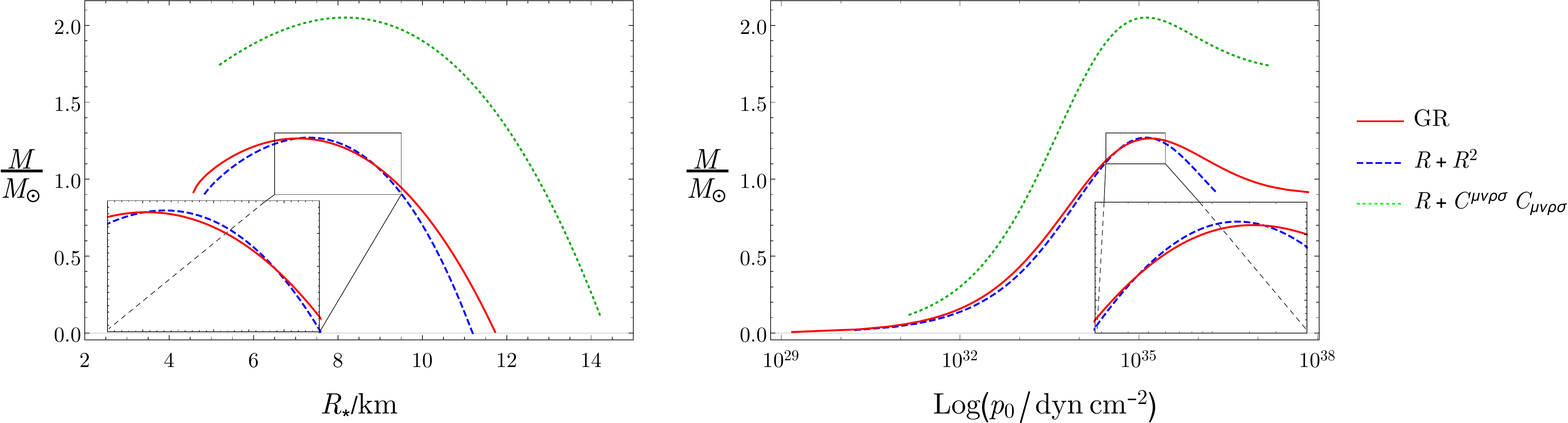}
\caption{ADM mass relations with radius and central pressure for GR, $R+R^2$ and $R+C_{\mu\nu\rho\sigma}C^{\mu\nu\rho\sigma}$.}
\label{mrhoxmactions}
\end{figure}

\begin{figure}[t]
\centering
\includegraphics[width=\textwidth]{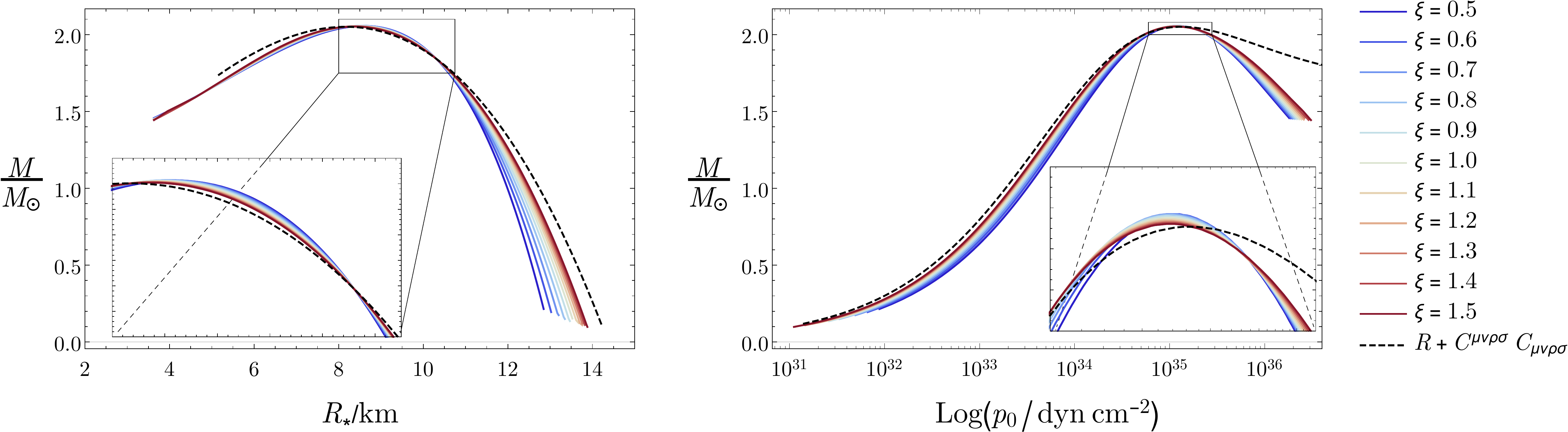}
\caption{ADM mass relations with radius and central pressure varying $\xi$, and the reference $R+C_{\mu\nu\rho\sigma}C^{\mu\nu\rho\sigma}$ solutions.}
\label{mrhoxm}
\end{figure}

While the impact of the massive modes on star structures might be interesting from a theoretical point of view, what is more relevant from an astrophysical perspective, and essential for a complete theoretical description, is their impact on the free parameters of the solutions. In figure \ref{mrhoxmactions} and \ref{mrhoxm} we show the relations of the asymptotic ADM mass $M$ with the star radius and central pressure, for the restricted theories and for the full quadratic theory respectively. In figure \ref{mrhoxmactions} the softening effect of the Weyl term is manifest, with the same pressure being able to sustain greater stellar masses, and with the same mass being bounded in larger volumes. We also note that this softening is quite impressive, with an increase in the maximum mass of a factor greater than 1.5. From both figure \ref{mrhoxmactions} and \ref{mrhoxm}, we see that the $R^2$ term does not always leads to a strengthening of gravity, but there is a trend inversion for higher masses, where the scalar contribution seams to weaken the gravitational interaction, resulting in an increase of the maximum mass. From figure \ref{mrhoxm} it also seems that there is a confirmation that the full quadratic theory is mainly affected by the Weyl term, with the scalar mode slightly modifying the parameters found in the $R+C_{\mu\nu\rho\sigma}C^{\mu\nu\rho\sigma}$ theory.


\begin{figure}[t]
\includegraphics[width=\textwidth]{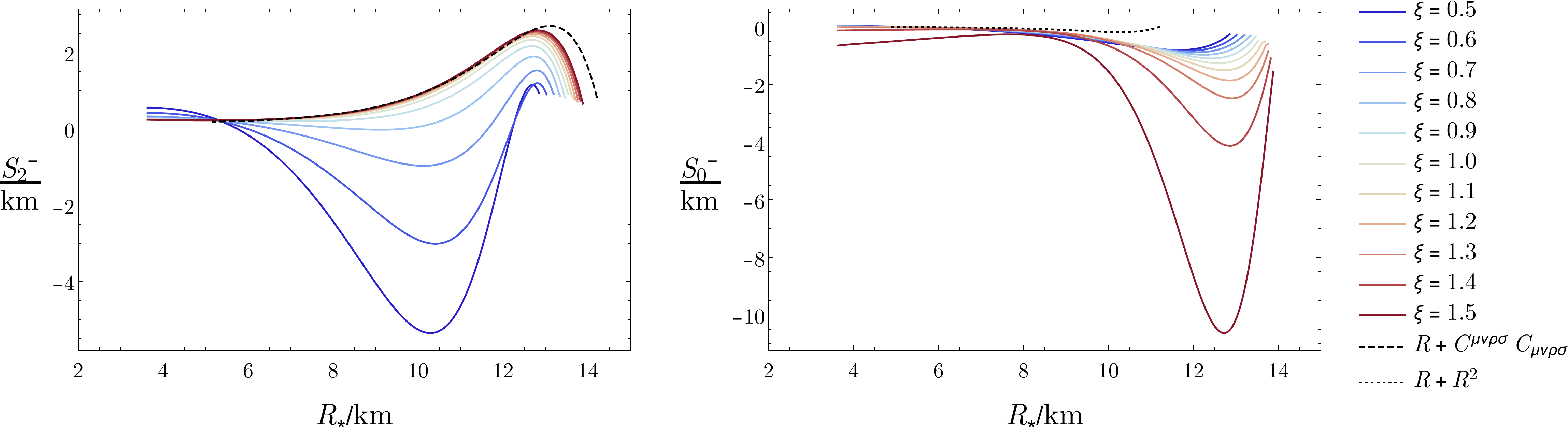}
\caption{$S_2^-$ and $S_0^-$ relation with the radius varying $\xi$, and the reference $R+C_{\mu\nu\rho\sigma}C^{\mu\nu\rho\sigma}$ and $R+R^2$ solutions.}
\label{s2s0xm}
\end{figure}

Even if a complete formal description of such effects has yet to be found, we can still associate these behaviours to the nature of the two massive modes. The Weyl term acts through a particle with negative energy states, therefore it is sensible to expect a repulsive contribution to gravity. The $R^2$ term, instead, acts through a massive scalar, that has a more familiar role in gravity. In particular we expect an attractive contribution to the interaction, but also an increase in stability, considering that the scalar field is distributed also outside the star. It is then sensible to have a strengthening of the gravitational interaction, but also a greater maximum mass. The role of the massive modes can be inferred also from the behaviour of the Yukawa parameters $S_2^-$ and $S_0^-$ shown in figure \ref{s2s0xm}. The scalar sector has somehow a predictable trend: it is always attractive, and the associated charge $S_0^-$ grows in order to contrast the decrease in range. The tensorial sector, on the contrary, can be both attractive and repulsive, as already seen in the black hole case \citep{Bonanno:2019rsq}. In particular for values of $\xi>1$ the contribution is always repulsive and the precise value has very little impact on the associated charge $S_2^-$, while for values $\xi<1$ it rapidly becomes attractive for most of the solutions. However, we have to remember that in this case the range of the scalar contribution is larger than the tensorial one, and then we can associate the attractive behaviour of the Weyl term to the presence of massive scalar particle outside the effective volume where the majority of the repulsive particles are present.
As final remark, we can differentiate the dependence of the solutions of the full quadratic theory from $\xi=m_0/m_2$ in three classes:
\begin{itemize}
\item $m_0>m_2$ the two Yukawa terms are competing, one being attractive and the other repulsive, and the scalar charge is larger than the tensorial one;
\item $m_0\sim m_2$ the two Yukawa terms are competing, one being attractive and the other repulsive, and the scalar charge is of the same order than the tensorial one;
\item $m_0<m_2$ the two Yukawa terms are both attractive, and the scalar charge is smaller than the tensorial one.
\end{itemize}
We would like to emphasize that a correction given by two competing and comparable Yukawa terms is more likely to be subject to a constraint different from the one found in \citep{Kapner:2006si,Giacchini:2016nta}. For the discussion on mass definitions we then opted for the value $\xi=1.1$, in order to have competing contributions of the same order and to avoid the fine tuned value $\xi=1$ for which some terms in the e.o.m. go to zero.

\subsubsection{Impact of the free parameters on the solutions}

In all the previous discussion we left unchanged the mass of one of the massive modes, or in other words the value of either $\alpha$ or $\beta$ in the action (\ref{action}). This approach is useful to emphasize the relative effects of the $R^2$ and Weyl term, but does not explain in detail the global effect of the quadratic terms. In this subsection we want to give a flavour of the impact of the free parameters, leaving a complete description to future work. We are indeed conscious that differences in the internal energy densities and pressures would have important consequences while using a realistic equation of state, but at the present time we limit ourselves to present the impact of the free parameters only at qualitative level using the polytropic e.o.s. In table \ref{table1} we show the variation of the maximum mass, the radius and central pressure for stars with mass equal to the GR maximum, in a similar way to what have been done in \citep{Astashenok:2017dpo}. The parameter $\alpha$, or $\beta$ for the $R+R^2$ theory, is in units of the value $\alpha_0$ (or $\beta_0$) for which the intrinsic length scale is equal to the Sun Schwarzschild radius. We confirm once again that the Weyl term has an impressive impact on the solutions, with both masses and radii increasing significantly as the value of $\alpha$ is increased; the $R^2$ term, instead, only slightly modifies the GR and the Einstein-Weyl solutions. We have also a confirmation of the softening effect of the Weyl term, having a decrease in the central pressure, and of the trend inversion of the strengthening effect of the $R^2$ term close to the maximum mass. We note, however, that the great differences in the internal pressures for different values of $\beta$ might be much more relevant while considering realistic equation of state.
\begin{table}[t]
\centering
 \begin{tabular}{|c|c|c|c|c|}
 \hline
Theory & $\alpha/\alpha_0$ or $\beta/\beta_0$ & $\Delta M_{max}/M_{\odot}$ & $\Delta R_{max}/km$ & $\Delta\,p_{c,max}/dyn\, cm^{-2}$ \\
 \hline
 \hline
 \multirow{5}{*}{$R+R^2$} & 1/2 & 0.001 & 0.350 & -0.969$\cdot 10^{35}$ \\ 
 & 1 & 0.005 & 0.539 & -0.544$\cdot 10^{35}$\\ 
 & 2 & 0.009 & 0.665 & 0.314$\cdot 10^{35}$\\ 
 & 5 & 0.013 & 0.770 & 2.897$\cdot 10^{35}$\\ 
 & 10 & 0.015 & 0.816 & 7.218$\cdot 10^{35}$\\ 
 \hline
 \multirow{5}{*}{$R+C_{\mu\nu\rho\sigma}C^{\mu\nu\rho\sigma}$} & 1/2 & 0.472 & 3.733 & -1.502$\cdot 10^{35}$\\ 
 & 1 & 0.786 & 5.112 & -1.505$\cdot 10^{35}$\\ 
 & 2 & 1.256 & 6.915 & -1.512$\cdot 10^{35}$\\ 
 & 5 & 2.225 & 10.177 & -1.524$\cdot 10^{35}$\\ 
 & 10 & 3.341 & 13.585 & -1.531$\cdot 10^{35}$\\ 
 \hline
 \multirow{5}{*}{Full quadratic, $\xi=0.5$} & 1/2 & 0.477 & 3.489 & -1.488$\cdot 10^{35}$\\ 
 & 1 & 0.791 & 4.549 & -1.481$\cdot 10^{35}$\\ 
 & 2 & 1.258 & 5.849 & -1.480$\cdot 10^{35}$\\ 
 & 5 & 2.209 & 8.004 & -1.484$\cdot 10^{35}$\\ 
 & 10 & 3.292 & 10.030 & -1.490$\cdot 10^{35}$\\ 
 \hline
 \multirow{5}{*}{Full quadratic, $\xi=1.0$} & 1/2 & 0.472 & 3.615 & -1.496$\cdot 10^{35}$\\ 
 & 1 & 0.790 & 4.808 & -1.495$\cdot 10^{35}$\\ 
 & 2 & 1.267 & 6.261 & -1.496$\cdot 10^{35}$\\ 
 & 5 & 2.243 & 8.644 & -1.501$\cdot 10^{35}$\\ 
 & 10 & 3.357 & 10.853 & -1.506$\cdot 10^{35}$\\ 
 \hline
 \multirow{5}{*}{Full quadratic, $\xi=1.5$} & 1/2 & 0.474 & 3.668 & -1.499$\cdot 10^{35}$\\ 
 & 1 & 0.787 & 4.936 & -1.450$\cdot 10^{35}$\\ 
 & 2 & 1.264 & 6.512 & -1.504$\cdot 10^{35}$\\ 
 & 5 & 2.247 & 9.140 & -1.511$\cdot 10^{35}$\\ 
 & 10 & 3.374 & 11.608 & -1.517$\cdot 10^{35}$\\ 
 \hline
 \end{tabular}
 \caption{Impact of the free parameters on the solutions; the parameters are taken in units of the length scales discussed in subsection \ref{sec3su2}, $\Delta M_{max}$ is the difference in the maximum mass of the stars, $\Delta R_{max}$ and $\Delta\, p_{c,max}$ are the differences of the radius and central pressure for stars with mass equal to the GR maximum.}
 \label{table1}
\end{table}
In conclusion, the variation of the free parameters of the theory does not change the behaviour of the solutions of quadratic gravity, but, as could be expected, has an impact only on the scales of the deviation from General Relativity.

\subsection{Discrepancies in mass definitions as a possible deviation from General Relativity}\label{massesdef}
\begin{figure}[t]
\centering
\includegraphics[width=\textwidth]{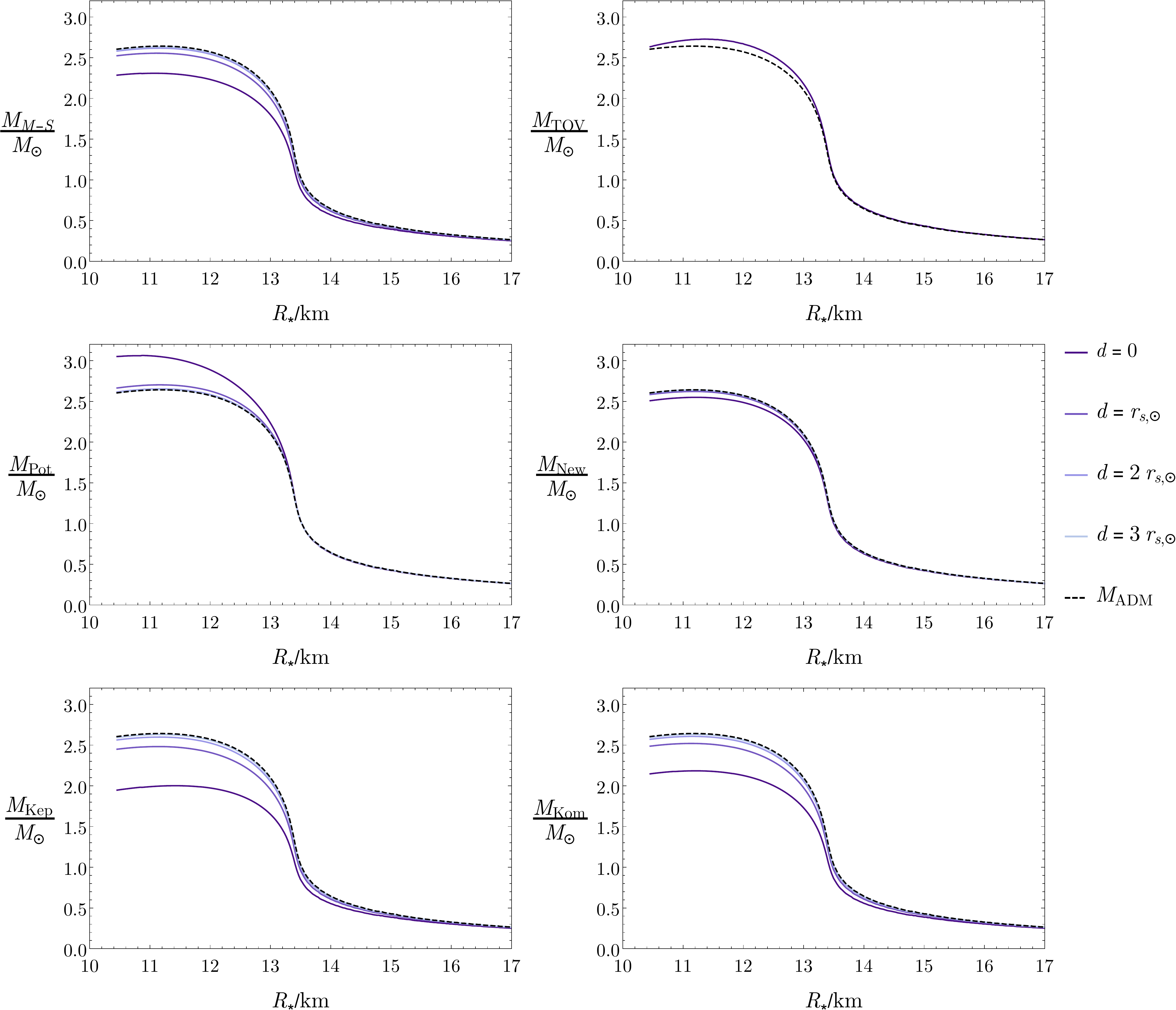}
\caption{Mass-radius relations for different mass definitions, evaluated at different distances from the star surface.}
\label{mr}
\end{figure}

We now move on the analysis of mass definitions in quadratic theories of gravity.
%, continuing the discussion of \citep{Resco:2016upv,Astashenok:2017dpo,Sbisa:2019mae}. 
The first point we want to stress is the importance of characterizing the external field in such theories. With the assurance of having the Schwarzschild metric on the outside, standard TOV solving methods in General Relativity focus only on the interior, and star masses are defined by the value of some function, namely either the Misner-Sharp (\ref{mhay}), the TOV (\ref{mtov}) or the Newtonian (\ref{mnew}) mass definition, at the star surface. Earlier works in quadratic theories of gravity used similar procedures \citep{Arapoglu:2010rz,Orellana:2013gn,Astashenok:2013vza,Capozziello:2015yza,Santos:2011ye,Deliduman:2011nw}, opening the possibility to ambiguities in the definition of mass, as noted by \citep{Sbisa:2019mae}. More recent works agree on the need of using asymptotic limits in order to extract information on neutron star masses avoiding discrepancies \citep{Yazadjiev:2014cza,Resco:2016upv,Astashenok:2017dpo}, however considering only the asymptotic limit may discard some relevant information, and both approaches come with consequences.

The characterization of the external field allows us to have no ambiguities and to lose no information. In figure \ref{mr} we show the mass-radius relations for the mass definitions of subsection \ref{massdef}, evaluated at different distances from the star surface $d=r-R_*$. Each definition has its specific mass-radius relation, and its specific dependence from the distance $d$, although it is a general feature that they coincide with the asymptotic limit after few solar Schwarzschild radius from the surface, with the exception of the TOV mass that is identically identified by its value at the star surface.

The large variety in the possible mass-radius relations of the same class of solutions leaves us with two possible approaches to the interpretation of mass in quadratic gravity:
\begin{itemize}
\item the only meaningful definition of mass is the asymptotic one $M$, and in order to completely define the gravitational field we have to measure the tensorial and scalar charges $S_2^-$ and $S_0^-$;
\item each mass definition describes a particular physical property of the star, and any star analysis has to consider carefully which aspect is under investigation. 
\end{itemize}
In agreement with \citep{Sbisa:2019mae}, 
we believe that the second approach is the most powerful one, given that it is less reliant on the underlying theory of gravity. In particular we can reverse our point of view and state that \emph{the presence of discrepancies between different mass definitions can be taken as a strong indication of departure from General Relativity}.

\begin{figure}[t]
\centering
\includegraphics[width=0.85\textwidth]{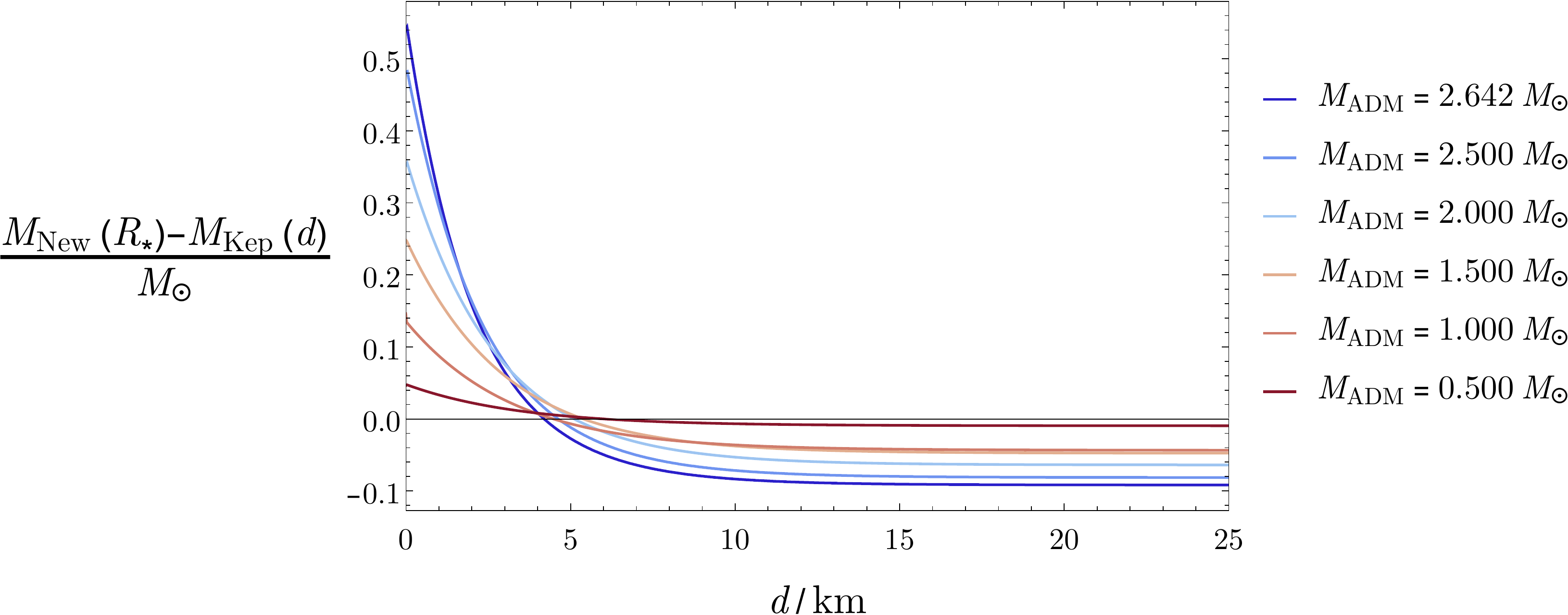}
\caption{Differences between masses measured with surface redshift or planetary transit, in function of the distance from the star surface, for stars with ADM masses in the range $[0.5-2.642]\,M_{\odot}$.}
\label{diffmass}
\end{figure}

We would like to emphasize this concept already stated in \citep{Sbisa:2019mae,Resco:2016upv}, and explicited in \citep{Astashenok:2017dpo}, and to consider the example of a neutron star with an orbiting planet: we can estimate the mass of this star from the planetary transit of its satellite using the third Kepler's law, and also from the redshift of photons emitted at the surface and measured at infinity. For the former measurement we have to use the Keplerian definition (\ref{mkep}) in function of the distance of the planet from the star, for the latter we have to use the Newtonian definition (\ref{mnew}) evaluated at the star surface. In figure \ref{diffmass} we show  the difference between the two measurements for the case of six different realistic stars with asymptotic ADM masses between $[0.5-2.642]\,M_{\odot}$.
%As a test to verify quadratic gravity this experiment should match the precise numerical values we found, %but we have to remember that we are assuming:
%\begin{itemize}
%\item[-] the scales of the solutions;
%\item[-] the ratio of the masses $\xi=m_0/m_2$;
%\item[-] the equation of state;
%\end{itemize}
%and then our numerical predictions are not reliable. 
%As a test to falsify General Relativity, however, the relevant result is that we are predicting non-zero %and macroscopical values, and the test is well-posed.
Albeit the large range of possibilities for the masses $m_2$ and $m_0$ might render problematic in practice to strongly constraint the 
predictions of the quadratic theory, it is also clear that the ambiguity in the definitions of mass in these theories can  be conveniently exploited to define deviations from classical General Relativity.

\section{Conclusions}\label{sec5}
In this paper we studied self gravitating fluid solutions of quadratic gravity in their entirety. In particular, thanks to an implementation of the shooting method, we linked the effects of quadratic curvature terms on the fluid to explicit Yukawa corrections of the external Newtonian potential. The inclusion of the Weyl squared term determines a major softening of gravity, while the $R^2$ term has a strengthening effect on the interaction, and a stabilizing effect on the solutions. The full quadratic theory is mainly affected by the presence of the Weyl term, with significant modifications due to the scalar sector only in the limit of low mass stars and of scalar particles lighter than tensorial ones. The global scales of the deviation from GR, instead, are completely determined by the value of one of the two free parameters in the action, which are only mildly constrained. The relative weights of the masses of the two Yukawa particles determine also an either attractive or repulsive contribution of the Weyl term, while the scalar contribution is always attractive. We believe that the presence of competing Yukawa corrections to the gravitational potential might be relevant for the constraints on the masses of these particles.
In the last subsection we analyzed different mass-radius relations for realistic neutron stars, taking into account different viable mass definitions evaluated at different distances from the stars. The large variety in the obtained relations, together with the additional variation due to the free parameters of the theory and the equation of state, suggests that quadratic gravity is a more effective tool if used to look for deviations from General Relativity.  It would be important to extend our analysis to
the study of the stability of the solutions as a function of the $\xi=m_0/m_2$ parameter and we plan to address this problem in future. 

\paragraph{Acknowledgements.} We would like to thank Alessia Platania and Massimiliano Rinaldi for their useful advice and stimulating discussions. This work has been partially supported by the INFN grant FLAG and the TIFPA - Trento Institute for Fundamental Physics and Applications.

\newpage

\appendix

\section{Full form of the equations of motions}\label{appendixA}

We present here for the sake of completeness the full form of the equations (\ref{eom3})
\begin{dmath}
\mathcal{G}_{rr}-\frac{1}{2}T_{rr}=\frac{1}{24 r^4 f(r) h(r)^4}\Bigg(-2 r^3 f(r) h(r) h'(r)^2 \big(3 r (\alpha +3 \beta ) f'(r) h'(r)+2 f(r) \big(3 r (\alpha -3 \beta ) h''(r)+(5 \alpha +12 \beta ) h'(r)\big)\big)-2 r h(r)^3 \Big(r^2 f'(r) \big(9 \beta  r
   f''(r) h'(r)+f'(r) \big(27 \beta  r h''(r)-2 (\alpha -12 \beta ) h'(r)\big)\big)+2 r f(r) \Big(r f'(r) \Big(9 \beta  r h^{(3)}(r)+(4 \alpha +42 \beta ) h''(r)\Big)+h'(r) \big(4 r (\alpha +6
   \beta ) f''(r)+2 (\alpha +15 \beta ) f'(r)-6 \gamma  r\big)\Big)-8 (\alpha +6 \beta ) f(r)^2 \Big(2 h'(r)-r \big(r h^{(3)}(r)+2 h''(r)\big)\Big)\Big)+r^2 h(r)^2 \big(-r^2 (\alpha -39 \beta
   ) f'(r)^2 h'(r)^2+4 r f(r) h'(r) \big(r (\alpha -3 \beta ) f''(r) h'(r)+2 (\alpha +6 \beta ) f'(r) \big(r h''(r)+2 h'(r)\big)\big)-4 f(r)^2 \big(r^2 (\alpha -3 \beta ) h''(r)^2+(\alpha -48 \beta
   ) h'(r)^2-2 r h'(r) \big(r (\alpha -3 \beta ) h^{(3)}(r)+6 (\alpha +3 \beta ) h''(r)\big)\big)\big)+4 h(r)^4 \big(4 \alpha -12 \beta-\alpha  r^2 f'(r)^2+12 \beta  r^2 f'(r)^2-36 \beta  r
   f'(r)-18 \beta  r^3 f'(r) f''(r)+f(r) \big(-72 \beta +4 r^2 (\alpha -12 \beta ) f''(r)+36 \beta  r f'(r)+6 \gamma  r^2\big)-4 (\alpha -21 \beta ) f(r)^2-3 r^4 p(r)-6 \gamma  r^2\big)+7 r^4 (\alpha
   -3 \beta ) f(r)^2 h'(r)^4\Bigg)=0;
\end{dmath}
\begin{dmath*}
\mathcal{G}_{tt}+X(r)\partial_r\big(\mathcal{G}_{rr}-\frac{1}{2}T_{rr}\big) +Y(r)\big(\mathcal{G}_{rr}-\frac{1}{2}T_{rr}\big)-\frac{1}{2}T_{tt} = \frac{1}{8
   r^4 h(r)^3 \big(9 r \beta  h(r) f'(r)+f(r) \big(4 (\alpha +6 \beta ) h(r)-2 r (\alpha -3 \beta ) h'(r)\big)\big)^2}\Bigg(6 \beta  \big(128 \big(5 \alpha ^2+33 \beta  \alpha +18 \beta ^2\big) h(r)^6+32 r \big(-25 \alpha ^2+15 \beta  \alpha +18 \beta ^2\big) h'(r) h(r)^5+16 r^2 \big(\big(7 \alpha ^2-132 \beta 
   \alpha -153 \beta ^2\big) h'(r)^2-4 r \big(7 \alpha ^2+48 \beta  \alpha +36 \beta ^2\big) h''(r) h'(r)+2 r^2 \big(\alpha ^2+3 \beta  \alpha -18 \beta ^2\big) h''(r)^2\big) h(r)^4-8 r^3 h'(r)
   \big(\big(-43 \alpha ^2-318 \beta  \alpha -36 \beta ^2\big) h'(r)^2-24 r \alpha  (\alpha -3 \beta ) h''(r) h'(r)+r^2 \big(5 \alpha ^2-21 \beta  \alpha +18 \beta ^2\big) h''(r)^2\big) h(r)^3+4
   r^4 h'(r)^2 \big(-27 \alpha  (\alpha -5 \beta ) h'(r)^2+2 r \big(5 \alpha ^2-21 \beta  \alpha +18 \beta ^2\big) h''(r) h'(r)+r^2 (\alpha -3 \beta )^2 h''(r)^2\big) h(r)^2-2 r^5 (\alpha -3 \beta )
   h'(r)^4 \big((11 \alpha -6 \beta ) h'(r)+2 r (\alpha -3 \beta ) h''(r)\big) h(r)+r^6 (\alpha -3 \beta )^2 h'(r)^6\big) f(r)^4+h(r) \big(-3 \beta  \big(8 \alpha ^2-39 \beta  \alpha +45 \beta
   ^2\big) f'(r) h'(r)^5 r^6-12 \beta  h(r) h'(r)^3 \big(r h'(r) f''(r) (\alpha -3 \beta )^2+f'(r) \big(\big(2 \alpha ^2-39 \beta  \alpha +18 \beta ^2\big) h'(r)-3 r \big(2 \alpha ^2-9 \beta 
   \alpha +9 \beta ^2\big) h''(r)\big)\big) r^5+4 h(r)^2 h'(r) \big(-3 \beta  \big(4 \alpha ^2-15 \beta  \alpha +9 \beta ^2\big) f'(r) h''(r)^2 r^3+12 (\alpha -3 \beta ) \beta  h'(r)
   \big((\alpha -12 \beta ) f'(r)+r (\alpha -3 \beta ) f''(r)\big) h''(r) r^2+h'(r)^2 \big(3 r \beta  \big(124 \alpha ^2+327 \beta  \alpha +252 \beta ^2\big) f'(r)+4 (\alpha -3 \beta ) \big(-2
   \alpha  \gamma  r^2-3 \beta  \gamma  r^2+18 \alpha  \beta  f''(r) r^2-36 \beta ^2+12 \alpha  \beta \big)\big)\big) r^3-16 h(r)^3 \big(3 (\alpha -3 \beta ) \beta  \big(r (\alpha -3 \beta )
   f''(r)-18 \beta  f'(r)\big) h''(r)^2 r^3+2 \big(3 \beta  \big(8 \alpha ^2+69 \beta  \alpha +126 \beta ^2\big) f'(r)-2 r (\alpha -3 \beta )^2 \gamma \big) h'(r) h''(r) r^2+3 h'(r)^2 \big(-36
   \beta ^3-84 \alpha  \beta ^2-9 r^2 \gamma  \beta ^2+32 \alpha ^2 \beta -3 r^2 \alpha  \gamma  \beta +2 r \big(32 \alpha ^2+231 \beta  \alpha +153 \beta ^2\big) f'(r) \beta +r^2 \big(41 \alpha ^2-57
      \end{dmath*}
\begin{dmath}
   \beta  \alpha -36 \beta ^2\big) f''(r) \beta -4 r^2 \alpha ^2 \gamma \big)\big) r^2-16 h(r)^4 \big(8 (\alpha +6 \beta ) \big(r (\alpha -3 \beta ) \gamma -9 \alpha  \beta  f'(r)\big) h''(r)
   r^2+h'(r) \big(-216 \alpha  \beta ^2 f^{(3)}(r) r^3+72 \alpha ^2 \beta  f^{(3)}(r) r^3+8 \alpha ^2 \gamma  r^2-252 \beta ^2 \gamma  r^2+6 \alpha  \beta  \gamma  r^2-12 \beta  \big(11 \alpha ^2+42
   \beta  \alpha +18 \beta ^2\big) f''(r) r^2-3 \beta  \big(76 \alpha ^2-141 \beta  \alpha +225 \beta ^2\big) f'(r) r+432 \beta ^3-72 \alpha  \beta ^2-240 \alpha ^2 \beta \big)\big) r+64 h(r)^5
   \big(3 \beta  \big(7 \alpha ^2+84 \beta  \alpha +9 \beta ^2\big) f''(r) r^2+6 \beta  \big(-16 \alpha ^2-39 \beta  \alpha +99 \beta ^2\big) f'(r) r+2 (\alpha +6 \beta ) \big(18 \alpha  \beta 
   f^{(3)}(r) r^3-(2 \alpha +3 \beta ) \big(\gamma  r^2+12 \beta \big)\big)\big)\big) f(r)^3+h(r)^2 \big(9 \beta  \big(4 \alpha ^2-15 \beta  \alpha +9 \beta ^2\big) f'(r)^2 h'(r)^4 r^6+12
   \beta  h(r) f'(r) h'(r)^2 \big(2 r h'(r) f''(r) (\alpha -3 \beta )^2+f'(r) \big(\big(11 \alpha ^2+6 \beta  \alpha +126 \beta ^2\big) h'(r)-6 r \alpha  (\alpha -3 \beta ) h''(r)\big)\big) r^5-4
   h(r)^2 \big(3 \beta  \big(-2 \alpha ^2+3 \beta  \alpha +9 \beta ^2\big) f'(r)^2 h''(r)^2 r^4+12 \beta  f'(r) h'(r) \big(r f''(r) (\alpha -3 \beta )^2+\big(2 \alpha ^2-3 \beta  \alpha +72 \beta
   ^2\big) f'(r)\big) h''(r) r^3+h'(r)^2 \big(2 \big(2 \alpha ^2-21 \beta  \alpha +45 \beta ^2\big) p(r) r^4+4 (\alpha -3 \beta )^2 \rho (r) r^4-8 \alpha ^2 \gamma  f'(r) r^3-18 \beta ^2 \gamma 
   f'(r) r^3-24 \alpha  \beta  \gamma  f'(r) r^3-216 \beta ^3 f'(r) f''(r) r^3-126 \alpha  \beta ^2 f'(r) f''(r) r^3+120 \alpha ^2 \beta  f'(r) f''(r) r^3+1404 \beta ^3 f'(r)^2 r^2+2493 \alpha  \beta ^2
   f'(r)^2 r^2+228 \alpha ^2 \beta  f'(r)^2 r^2+108 \beta ^2 \gamma  r^2-36 \alpha  \beta  \gamma  r^2-216 \beta ^3 f'(r) r-72 \alpha  \beta ^2 f'(r) r+48 \alpha ^2 \beta  f'(r) r+216 \beta ^3-144 \alpha 
   \beta ^2+24 \alpha ^2 \beta \big)\big) r^2-8 h(r)^3 \big(h'(r) \big(4 \alpha ^2 p'(r) r^5+36 \beta ^2 p'(r) r^5-24 \alpha  \beta  p'(r) r^5-108 \alpha  \beta ^2 f''(r)^2 r^4+36 \alpha ^2 \beta 
   f''(r)^2 r^4+2 \big(4 \alpha ^2-15 \beta  \alpha +90 \beta ^2\big) p(r) r^4-8 \big(\alpha ^2+3 \beta  \alpha -18 \beta ^2\big) \rho (r) r^4+108 \beta ^2 \gamma  f''(r) r^4-36 \alpha  \beta  \gamma
    f''(r) r^4-108 \alpha  \beta ^2 f'(r) f^{(3)}(r) r^4+36 \alpha ^2 \beta  f'(r) f^{(3)}(r) r^4-8 \alpha ^2 \gamma  f'(r) r^3-153 \beta ^2 \gamma  f'(r) r^3+48 \alpha  \beta  \gamma  f'(r) r^3-918 \alpha
    \beta ^2 f'(r) f''(r) r^3+36 \alpha ^2 \beta  f'(r) f''(r) r^3-432 \beta ^3 f'(r)^2 r^2-684 \alpha  \beta ^2 f'(r)^2 r^2-210 \alpha ^2 \beta  f'(r)^2 r^2+16 \alpha ^2 \gamma  r^2-72 \beta ^2 \gamma 
   r^2+84 \alpha  \beta  \gamma  r^2+432 \beta ^3 f''(r) r^2-288 \alpha  \beta ^2 f''(r) r^2+48 \alpha ^2 \beta  f''(r) r^2+1404 \beta ^3 f'(r) r-936 \alpha  \beta ^2 f'(r) r+48 \alpha ^2 \beta  f'(r)
   r-432 \beta ^3+504 \alpha  \beta ^2-120 \alpha ^2 \beta \big)-12 r^3 \beta  f'(r) \big(2 \big(2 \alpha ^2+24 \beta  \alpha -9 \beta ^2\big) f'(r)+r (\alpha -3 \beta ) \big((\alpha +6 \beta )
   f''(r)-3 \gamma \big)\big) h''(r)\big) r+16 h(r)^4 \big(4 \alpha ^2 p'(r) r^5-72 \beta ^2 p'(r) r^5+12 \alpha  \beta  p'(r) r^5-108 \alpha  \beta ^2 f''(r)^2 r^4+36 \alpha ^2 \beta  f''(r)^2
   r^4+12 \big(\alpha ^2+3 \beta  \alpha -18 \beta ^2\big) p(r) r^4-4 (\alpha +6 \beta )^2 \rho (r) r^4-54 \beta ^2 \gamma  f''(r) r^4+18 \alpha  \beta  \gamma  f''(r) r^4+540 \alpha  \beta ^2 f'(r)
   f^{(3)}(r) r^4+36 \alpha ^2 \beta  f'(r) f^{(3)}(r) r^4-16 \alpha ^2 \gamma  f'(r) r^3-360 \beta ^2 \gamma  f'(r) r^3-156 \alpha  \beta  \gamma  f'(r) r^3+864 \alpha  \beta ^2 f'(r) f''(r) r^3+144
   \alpha ^2 \beta  f'(r) f''(r) r^3+1971 \beta ^3 f'(r)^2 r^2-1233 \alpha  \beta ^2 f'(r)^2 r^2-78 \alpha ^2 \beta  f'(r)^2 r^2+16 \alpha ^2 \gamma  r^2+144 \beta ^2 \gamma  r^2+120 \alpha  \beta  \gamma 
   r^2-216 \beta ^3 f''(r) r^2-72 \alpha  \beta ^2 f''(r) r^2+48 \alpha ^2 \beta  f''(r) r^2-3024 \beta ^3 f'(r) r-576 \alpha  \beta ^2 f'(r) r+96 \alpha ^2 \beta  f'(r) r+864 \beta ^3-144 \alpha  \beta
   ^2-48 \alpha ^2 \beta \big)\big) f(r)^2+3 r \beta  h(r)^3 \big(\big(-8 \alpha ^2
   +21 \beta  \alpha + 9 \beta ^2\big) f'(r)^3 h'(r)^3 r^5 + 4 h(r) f'(r)^2 h'(r) \big(f'(r) \big(\big(4 \alpha^2 - 105 \beta  \alpha - 126 \beta ^2\big) h'(r) + r \big(2 \alpha ^2 - 3 \beta  \alpha - 9 \beta ^2\big) h''(r)\big) - r (\alpha - 3 \beta )^2 h'(r) f''(r)\big) r^4 + 4 h(r)^2 f'(r) \big(2 \big(-2 \alpha^2 + 21 \beta  \alpha + 36 \beta ^2\big) f'(r)^2 h''(r) r^3 + h'(r) \big(-27 \beta  p(r) r^4 + 12 (\alpha - 3 \beta ) \rho (r) r^4 - 12 \alpha  \gamma  f'(r) r^3 + 36 \beta  \gamma  f'(r) r^3 + 4 \alpha^2 f'(r) f''(r) r^3-72 \beta ^2 f'(r) f''(r) r^3 + 66 \alpha  \beta  f'(r) f''(r) r^3 + 4 \alpha ^2 f'(r)^2 r^2 + 36 \beta ^2 f'(r)^2 r^2 + 273 \alpha  \beta  f'(r)^2 r^2 - 24 \alpha  \gamma  r^2 + 18 \beta  \gamma  r^2 + 16 \alpha ^2 + 36 \beta ^2 - 60 \alpha  \beta \big)\big) r - 8 h(r)^3 \big(4 r^2 \big(\alpha ^2 + 30 \beta  \alpha - 99 \beta^2\big) f'(r)^3 + \big(2 \big(\alpha ^2 - 150 \beta  \alpha + 36 \beta ^2\big) f''(r) r^3 + 3 \big(-18 \alpha  \beta  f^{(3)}(r) r^3 + 14 \alpha  \gamma  r^2 + 39 \beta  \gamma  r^2 + 252 \beta ^2 - 48 \alpha  \beta \big) r\big) f'(r)^2 + 6 \big(-\alpha  p'(r) r^5 + 3 \beta  p'(r) r^5 - 2 (\alpha - 3 \beta ) p(r) r^4 + 2 (\alpha + 6 \beta ) \rho (r) r^4 - 4 \alpha  \gamma  r^2 - 24 \beta  \gamma  r^2 - 72 \beta ^2 + 24 \alpha  \beta \big) f'(r) + 2 r (\alpha - 3 \beta ) \big(3 p(r) r^4 + 6 \gamma r^2 - 4 \alpha + 12 \beta \big) f''(r)\big)\big) f(r) + 3 r^2 \beta  h(r)^4 f'(r)^2 \big(\big(2 \alpha^2 - 3 \beta  \alpha - 9 \beta ^2\big) f'(r)^2 h'(r)^2 r^4 + 4 \big(-2 \alpha^2 + 21 \beta  \alpha + 36 \beta^2\big) h(r) f'(r)^2 h'(r) r^3 + 4 h(r)^2 \big(6 (\alpha - 3 \beta ) p(r) r^4 - 27 \beta  \rho (r) r^4 - 54 \beta  \gamma  f'(r) r^3 + 54 \alpha  \beta  f'(r) f''(r) r^3 + 2 \alpha ^2 f'(r)^2 r^2 + 180 \beta^2 f'(r)^2 r^2 - 39 \alpha  \beta  f'(r)^2 r^2 + 12 \alpha  \gamma  r^2 + 18 \beta  \gamma  r^2 - 216 \beta ^2 f'(r) r + 72 \alpha  \beta  f'(r) r - 8 \alpha^2 + 36 \beta^2 + 12 \alpha  \beta \big)\big)\Bigg) = 0.
\end{dmath}

\bibliographystyle{JHEP}
\providecommand{\href}[2]{#2}\begingroup\raggedright\endgroup

\end{document}